\documentclass{iauc}
\usepackage{graphicx}

\title[A review on success and problem of MOND]
{Modified Newtonian Dynamics: success and problem on globular cluster scale}

\author[HongSheng Zhao]   
{HongSheng Zhao} \affiliation{University of St Andrews, School of Physics and Astronomy, KY16 9SS, UK\footnote{PPARC Advanced Fellow (hz4@st-and.ac.uk)}\\
National Astronomical Observatories, Chinese Academy of
Sciences, Beijing, 100012, China\thanks{Overseas Outstanding Young Scholar}
\\[\affilskip]
}

\pubyear{2005}
\volume{198} 
\pagerange{1--6}
\date{20 June 2005}
\jname{Invited review for IAU Colloquium 198}
\editors{Helmut Jerjen and Bruno Binggeli eds.}

\def\beq{\begin{equation}}
\def\eeq{\end{equation}}
\def\bey{\begin{eqnarray}}
\def\eey{\end{eqnarray}}
\def\pc{\, {\rm pc} }

\def\Myr{\, {\rm Myr} }
\def\Gyr{\, {\rm Gyr} }
\def\msun{M_\odot}

\def\Lsun{L_\odot}
\def\lsun{L_\odot}
\def\kms{\, {\rm km \, s}^{-1} }
\def\vg{{\bf g}}
\def\vR{{\bf R}}
\def\grad{{\bf \nabla}}

\def\a0{$a_0$}

\begin{document}

\maketitle

\begin{abstract}
Many past attempts to kill MOND have only strengthened the theory.
Better data on galaxy velocity curves clearly favor MOND (without fine-tuning)
over cold dark matter.  The usual critism on the incompleteness
of classical MOND has spurred a Modified Relativity (MR) by Bekenstein.
After outlining cosmology and lensing in MOND, we review MOND on small scales.
We point out some potential problems of MOND in two-body relaxation and
tidal truncation.  We argue that the tidal field in any MOND-like
gravity theory predicts that the Roche lobe sizes of a binary
system are simply proportional to the binary baryonic mass ratio
to the power 1/3. An immediate application of this result is that
the tidal field and tidal truncation radii of million-star
globular clusters and million-star dwarf galaxies (e.g., the Milky
Way satellites NGC2419 and Carina) would be very similar because
of the one-to-one relation between gravity and baryon
distribution.  This prediction appears, however, inconsistent with
the fact that {\it all} globulars are truncated to much smaller
sizes than {\it all} dwarf galaxies. Whether tide is uniquely
determined by baryons can also be used to falsify any MOND-like
gravity theory, whether classical or relativistic.

\keywords{dark matter, galaxy dynamics}

\end{abstract}

\firstsection 
\section{Introduction}

A large amount of astronomical data suggests that the $GM/r^2$
Newtonian gravity from the baryonic material in galaxies fails to
explain the large accelerations implied by motions in spiral and
elliptical galaxies. However, Zwicky's proposal to
introduce unseen gravitating (dark) mass with a very flexible
density distribution wherever needed for the motions of galaxies
and stars has been troubled by a lack of experimental confirmation
of its elementary particle counterparts for more than 60
years. Evidence is also summarized in recent articles
about the difficulties facing dark matter (DM) theories on galaxy
scales (Spergel \& Steinhardt 2000, Ostriker \& Steinhardt 2003).
Surprisingly, an alternative gravity theory with Modified
Newtonian Dynamics (MOND, Milgrom 1983) has been doing very well
exactly where the dark matter theory is doing poorly. The
predictive power of this 20-year-old classical theory with
virtually no free parameters (Bekenstein \& Milgrom 1984) is recently highlighted by
the astonishingly good fits to contemporary kinematic data of a
wide variety of high and low surface brightness spiral and
elliptical galaxies; even the fine details of the ups and downs of
velocity curves are elegently reproduced without fine tuning of
the baryonic model (Sanders \& McGaugh 2002, Milgrom \& Sanders 2003). Originally
it was proposed that galaxy rotation curves could be fit by
motions in a modified gravity $\vg$, which is significantly
stronger than Newtonian gravity in the weak regime defined by a
gravitational field energy density \beq {|\vg|^2 \over 8\pi G }
\le {a_0^2 \over 8\pi G} \approx \rho_b(0) c^2, \eeq where $G$ is
the gravitational constant, and $\rho_b(0) c^2$ is the present
universal mean baryon energy density (commonly the weak regime is
defined as where gravity $g \le a_0 \sim 1.2\times 10^{-8} {\rm
cm}\sec^{-2}$). Far away from the baryons,
the modified gravity $\vg$ satisfies
\beq {\vg \over
\tilde{G}(g)} \approx {\vg_N \over G} \equiv - \sum_i {M_i \hat{\vR}_i \over
|R_i|^2} \eeq\label{farfield}
where $\vg_N$ is the vector sum of Newtonian gravity
(distance-inverse-squared attractive force) of all baryonic
particles with masses $M_i$ at distances $R_i$, where the
effective gravitational constant $\tilde{G}$ can be chosen as,
e.g., $ {G \over \tilde{G}(g)}  =  1 - \exp\left[-\left({g \over
a_0}\right)^\alpha\right], $ where $\alpha=1$ gives the MOND
gravity $\vg$. We get a MOND-like theory for any positive
$\alpha$. In these theories, the only matter that matters is the
luminous (baryonic) matter, and the gravitational field is a {\it
unique} function of the luminous matter distribution.

These alternative theories based on modifying the law of gravity
in a purely baryonic universe have been gaining ground rapidly.
Milgrom's empirical formula is a simplified treatment of a
classical theory with a curl-free gravitational field by
Bekenstein \& Milgrom (1984), which is now extendable to a
Modified Relativistic (MR) theory (Bekenstein 2004) which passes standard and
cosmological tests used to check General Relativity
(GR). This recent breakthrough by Bekenstein
transforms the empirical MOND formula to a falsifiable theory with
respectable theoretical underpinnings comparable almost to the
rival dark matter theory.   Here we examine various aspects of the MOND theory
after outlining cosmology and lensing in the relativistic MOND.

\subsection{Cosmology and Lensing}

Bekenstein (2004) has shown that MONDian cosmology and lensing follow the standard GR formulation.
The physical metric of the FRW universe in MOND is written as,
\beq
{g}_{\mu\nu} dx^\mu dx^\nu = c^2 dt^2 -
R(t)^2 \left[ d\chi^2 + f(\chi) ^2 \left(d\theta^2 + \sin^2\theta d\phi^2 \right) \right]
\eeq
where $f(\chi)=\sin\chi, \chi, \sinh \chi$ for closed, flat and open universe as in FRW GR,
and the physical scale factor is $R(t)=R_0/(1+z)$.  Note that we have removed
the tilde sign that Bekenstein uses to denote physical coordinates.
if we introduce the vaccum energy and neglect the scale field $\phi$ as in Bekenstein,
then the MONDian universe expands with a Hubble parameter $H(z(t))={d\ln R(t) \over dt}$
given approximately by
\beq
{H^2(z) \over H_0^2} \approx \Omega_m (1+z)^3 + \Omega_r (1+z)^4 +
\Omega_K (1+z)^2 + \Omega_\Lambda , \Omega_K \equiv 1- \Omega_m -\Omega_r - \Omega_\Lambda ,
\eeq
where $\Omega_r \sim 10^{-4} h^{-2}$ is normalised by the 2.7K CMB,
and $\Omega_m \sim 0.05 h^{-2}$ from all observable baryons,
and $\Omega_K$ is the curvature term.
The matter-radiation equality happens at redshift of about $\Omega_m / \Omega_r  \sim 1000$;
this is much lower than in CDM models.

To do lensing, Bekenstein shows (in eq. 116)
that for small perturbations of the metric by a potential $\Phi$, light travels on geodesic given by
\beq
ds^2 = \tilde{g}_{\mu\nu} dx^\mu dx^\nu \approx - (1+{2\Phi \over c^2}) c^2 dt^2 + (1-{2\Phi \over c^2}) dl^2 = 0,
\eeq
where $dl$ is the proper length.
Hence the light travel time
\beq
dt \approx (1 - {2\Phi \over c^2}) {dl \over c} ,
\eeq
same as in Einstein universe.
The gravititional potential satisfies approximately the MOND eq.
\beq
-\grad \cdot \left[ \mu({g \over a_0}) {\bf g} \right] = 4 \pi G \rho, \qquad {\bf g}=-\grad \Phi,
\eeq
where $\rho$ is the gravitational mass density of baryonic matter, and the $\mu$ is
the usual dimensionless function of the rescaled gravity ${g \over a_0}$ introduced by Milgrom.

The deflection for a light array of the closest approach (roughly
impact parmameter, physical length) $b$ is calculated by \beq
\alpha = \int {2 g_{\perp} d t \over c} \eeq where $g_{\perp} =
g(r) \cdot {b \over r}$ is the perpendicular component of the
gravity $g(r)$ at radius $r=\sqrt{(c t)^2+b^2}$, and $c t$ is the
path length along the line of sight counted from the point of
closest approach. Hence the deflection angle \beq\label{alphadt}
\alpha = \int_{-\infty}^\infty  {2 g(r)  d t \over c} {b \over
\sqrt{(c t)^2+b^2} }. \eeq This is essentially the same as eq.
(109) of Bekenstein.

\section{Two-body Relaxation Time in MOND}

As pointed by Ciotti \& Binney (2004), relaxation in MOND is a
non-trivial calculation because it is a long-range gravity with a
divergent potential at large distance. However, by and large, MOND
gravity is effectively a stronger-than-Newtonian gravity. This
suggests that we can rescale Newtonian gravity by a factor
${\tilde{G} \over G}$.  To balance this stronger gravity, a bound
object of certain size $r$ acquires a higher internal velocity
(than Newtonian case with pure baryons), hence shorter dynamical
time, and faster relaxation since the relaxation time is
proportional to dynamical time with \beq t_x \propto N t_{\rm dyn}
\sim {N r \over C V}, \eeq where $C \sim 7 \ln N \sim 50$ is a
large constant related to the Coulomb logarithm, and the dynamical
time \beq t_{\rm dyn}=\left({r \over V}\right) \sim 1\Myr {r \over
\pc}{1\kms \over V} \eeq where $r$ is the half-mass radius, and
$V$ is characteristic circular velocity. So for $N\sim 10^{4-5}$
globular cluster, it relaxs on time scales of 100-1000 dynamical
times. Typical dynamical time is of order Myr, hence a globular is
generally relaxed in a fraction of the Hubble time. In MOND the
circular velocity is generally given by \beq V = \left( G M a_0
\right)^{1/4}=3.5\kms\left( {M \over 10^4\msun} \right)^{1/4} \eeq
if in deep-MOND where $r \ge \sqrt{GM \over a_0}=3.5\pc \sqrt{M \over
10^4\msun}$.

Relaxation drives a globular to expand.  Star clusters might have
been born very compact, and in the
strong-gravity regime.  However, fast relaxation means that
globular clusters cannot stay small, will eventually expand and
enter the weak regime.
The relaxation time can be expressed in terms of some kind of
rate of expansion per relaxation time
\beq {r \over t_x} =  { 50 \pc \over 10\Gyr} {10^4 \over
N} \left({V \over 1\kms}\right), \qquad V=3.5\kms \left( {M \over
10^4\msun} \right)^{1/4} , \eeq where we have assumed $C \sim 7\ln
N \sim 50$, and have used the circular velocity in the MOND
regime. The above implies that a cluster should expand to a size
of $\sim 175$pc over a Hubble time in MOND.

\begin{figure}
\resizebox{9cm}{!}{\includegraphics{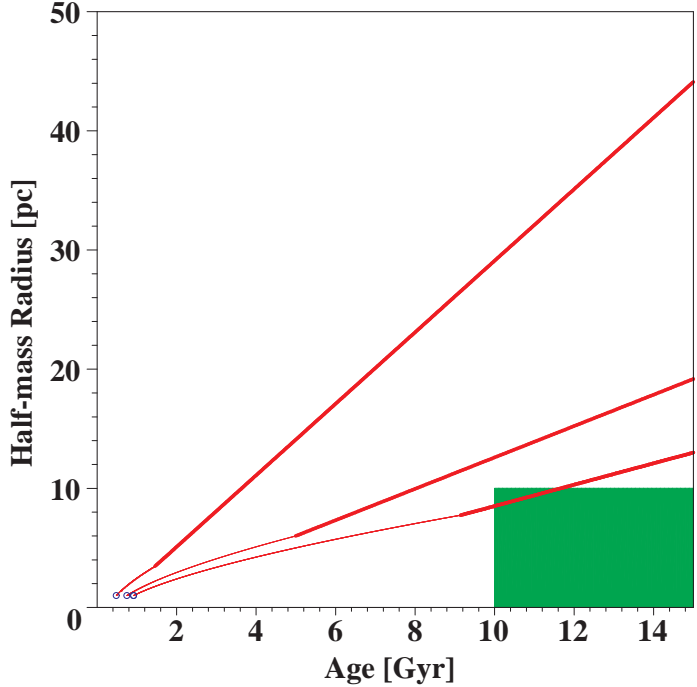}}
\caption{ shows the relaxation-driven expansion of the
half-mass radius of an isolated globular cluster
of $N=2 \times 10^4, 6 \times
10^4, 1 \times 10^5$ (curves from top to bottom) half-solar-mass
stars with a pre-core-collapse size of 1pc (marked by the blue
circles). The change from strong to weak gravity is indicated by
the change from thin lines to thick lines. Observed globulars
typically occupy the green shaded zone in the age vs. size plane.
}\label{rhage}
\end{figure}

To illustrate this, we give a few examples
in Fig.~\ref{rhage}.  Consider, for example, a hypothetical cluster
born fully isolated with $N=2 \times 10^4$ half-solar-mass stars
inside a half-mass radius of 1pc. The cluster is initially dense
and in the strong gravity regime. Rapid relaxation leads to core
collapse after about $t_0=0.4$ Gyrs where we take $t_0$ as $10$
times the initial relaxation time $t_x$. Afterwards the cluster
expands as $t^{2/3}$ in strong gravity till $r_h$ reaches about
3pc at about $t=1.4$ Gyr, after which the cluster enters the weak
gravity regime, $r_h$ expands linearly with time according to
MOND, and reaches a size $r_h=45$pc after a Hubble time.
Systems with larger N grows slower (Fig.~\ref{rhage}).
Nevertheless, clusters with $N=2\times 10^4$ to $1\times 10^5$
half-solar-mass stars all grow to a half-mass size $12-45$ pc,
too big to be consistent with observed globular clusters.

\section{Tidal truncations or Roche lobes of binary systems}

First we note that at large distances from the baryonic particles Gauss's
theorem predicts that a MOND-like gravity $\vg$ with an effective gravitational
constant $\tilde{G}(g)$ is approximately parallel to
the direction of the Newtonian gravity $\vg_N$ with the amplitude satisfying
eq.~(\ref{farfield}),
so the gravity $\vg$ is determined by the total mass of the baryons $\sum_i
M_i$ and the mean distance $r$, independent of the spatial distribution of the baryons.
This is true even in the rigorous classical counterpart (Bekenstein \& Milgrom 1984)
of the Bekenstein theory.
So, e.g., a dense globular cluster and a fluffy dwarf galaxy of
$10^5\lsun$ should have comparable baryonic mass, hence the similar
gravity field on scales of kpc, nearly independent of the internal
density profile.
Indeed MOND nicely accounts for the very small scatter of dynamical
properties of structures of similar baryonic distribution on galaxy scale
(Sanders McGaugh 2002, McGaugh 2000, 2004).  However,
this "nice feature" of baryonic gravity
also means that the dynamics of wide binary stars would be a scaled-down version of
the Antenna-like merging galaxies.

\begin{figure}
\resizebox{9cm}{!}{\includegraphics{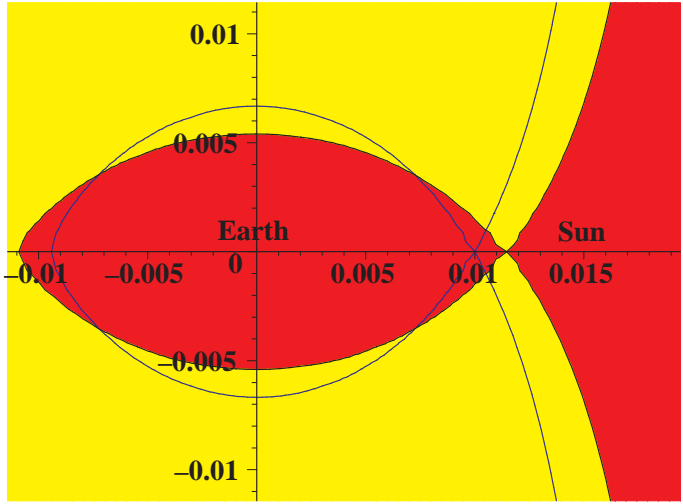}} \caption{ shows
the rescaled Roche lobes (contours of the effective potential) in
the equatorial plane of a hypothetical isolated Earth-Sun
binary mass ratio $3\times 10^{-6}$ in the strong gravity regime
(say with the separation 1AU, thin blue lines), and in the weak
gravity regime (say with separation 1pc, and $G\msun/1\pc^2 \sim
0.001a_0 \ll a_0$, shaded areas). The Earth is at origin and the
Sun is at rescaled unit length to the right. The inner Lagrangian
point is a saddle point between the Earth and the Sun, which is
slightly further away from the Earth in the deep-MOND regime than
in strong gravity regime. }\label{figroche0}
\end{figure}

Theoretically in a binary system, the Roche radius (or tidal
radius or virial radius when referring to galaxies sometimes) is
where the average density of the mass-losing satellite is
comparable to the average density of the mass-receiving host at
some point $R$ along the orbit of satellite, i.e.,
\beq\label{roche} {G m \over {4\pi \over 3} r_t^3} \approx k {G M
\over {4\pi \over 3} R^3}, \eeq with the prefactor $k$ ranging
from $1$ to $3$ in different definitions in the literature
(Binney \& Tremaine 1987);
e.g., $k=2$ according to von Hoerner (1957). This well-known tidal
criterion in Newtonian gravity is in fact a result of basic
dynamics, independent of the force law parameter $\tilde{G}(g)$
and applies to any gravity.  It is shown elsewhere (Zhao et al. 2005)
that the prefactor $k=2$ rigourously in the deep-MOND regime.
The shape of the Roche lobes is somewhat more squashed than
the Newtonian counterpart.  An example of Roche lobes is given
in Fig.~\ref{figroche0}.

\subsection{Application to satellite globular clusters and dwarf galaxies}

\begin{figure}
\resizebox{9cm}{!}{\includegraphics{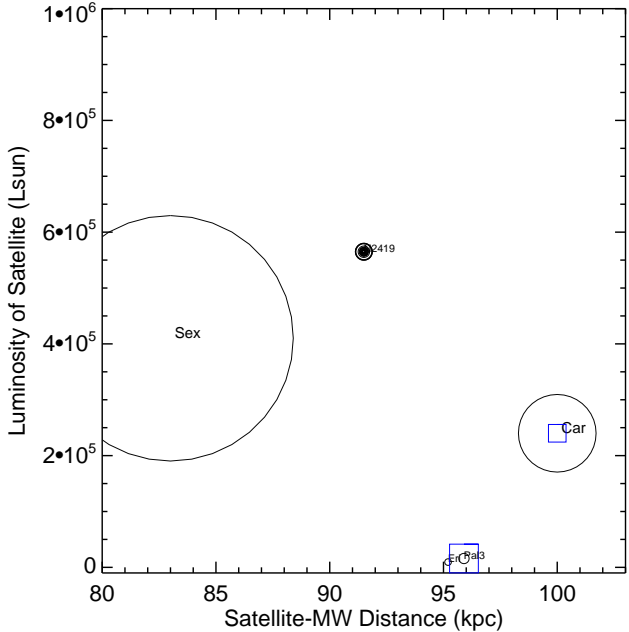}}
\caption{shows the
satellite distance vs. luminosity scatter diagram for the dwarf
galaxies (labeled with larger font) and globular clusters (labeled
with smaller font) orbiting in Milky Way's very outer halo (with
orbital periods about one quarter of a Hubble time). The sizes of
the symbols are scaled to the physical sizes of the satellites (as
a reference, the faint Eridanius cluster at the bottom is 100pc,
and Sextans dwarf is 3000pc).  The sizes of all globulars of the
Milky Way are all drawn as smaller rings inside NGC2419, which is
the largest, and most distant bright globular cluster. Predicted
sizes of the faint Pal3 cluster and Carina dwarf are also shown as
squares based on their proper motion inferred pericentres in a
$10^{11}\Lsun$ baryonic Milky Way.  }\label{satFig}
\end{figure}

The Roche criteria (eq.~\ref{roche}) predicts that objects should
have same sizes if they have similar baryonic content in similar
enviornment. This, however, is inconsistent with basic observation
data, and we see a large scatter for the truncation radii of outer
globular and a dozen dwarf galaxy satellites accelerating in the
tide and very weak gravity of a luminous host galaxy, like the
Milky Way (see Fig.~\ref{satFig}). So the observed limiting radii
of these satellites (Harris 1996, Mateo 1998) are no longer simply
and uniquely scaled with the binary baryonic mass ratio as
predicted by a MOND-like theory (eq.~\ref{roche}).

For example, NGC~2419, the largest cluster of the Milky Way, is as
distant as and 1-2 times as luminous as the $(2.5-4)\times
10^5\Lsun$ dwarfs Carina/Sextans, which has an equilibrium of
stars extending to 1000-3000pc radii at a distance of about 90kpc
from the Milky Way. A baryonic universe predicts that NGC~2419
should resist the tide of the Milky Way as well as the Sextans
dwarf does on 3000pc scale; Sextans shows no sign of being
disrupted (Grebel et al. 2000). Nevertheless, even this largest globular is
truncated on 200pc scale. Such large discrepancy does not go away
if we take into account of different orbits or mass of the
satellites, e.g., a baryonic universe would predict that the
cluster Pal3 (faint $10^4\lsun$ and presently near pericentre at
90 kpc, Dinescu et al. 1999) is bigger than the Carina dwarf (presently
near apocentre, Piatek et al. 2003), but in reality as shown in
Fig.~\ref{satFig}, Pal3 (Sohn et al. 2003) is much smaller than Carina
(Walcher et al. 2003). Note that orbits in MOND is not much different from
orbits in isothermal halo potentials (Read \& Moore 2005).

This example points out a generic problem general to galaxies with
resolved satellite globulars vs. dwarfs in any baryonic universe
including Bekenstein's cosmology: why there is a clear size gap
between 250pc-400pc and nature forms neither very big globulars
nor very small dwarfs despite a good mix of the baryonic mass and
enviornment variables (orbital phase and pericentric tide) of
these satellites. In a baryonic universe "globulars" and "dwarf
ellipticals" are practically similar assembly of stars described
by the King models: they are pressure-supported low-dispersion
gas-poor low-luminosity elliptical equilibria immersed in and
sculptured by the tidal field of a common host galaxy.

\subsection{Application: the sizes of the globular clusters in Fornax}

Could it be that the globulars were born small and stayed small?
It is plausible that globulars might have formed from denser and
cooler gas clouds than stars in some dwarfs. However, a very dense
core of a globular is theoretically unstable to stellar encounters
and must evolve rapidly, and on a Hubble time typically 10-100
percent of the initial mass of a star cluster are unbound due to
relaxation-driven stellar
evaporation (Johnstone 1993). Relaxation also
drives the half-light radius of a globular to expand as a power
law of the time $t^{2/3}$ if we assume globulars are born
infinitely compact and dense so we can neglect the delay for the
core to collapse.  Using eq. (1) and Fig. 1 of the numerical
simulations of (Giersz \& Heggie 1996), we estimate the radius containing 90
percent of the mass of a typical globular cluster of $10^5$
half-solar mass stars expands as $100{\rm pc}\left({t \over 10{\rm
Gyr}}\right)^{2/3}$.

Relaxation is even faster in MOND (Ciotti \& Binney 2004) because of the
stronger gravity, higher velocity dispersion and faster dynamical
time (Baumgardt et al. 2005).  A detailed calculation of the expansion rate
in MOND is given elsewhere (Zhao et al. 2005). Essentially  a globular
cluster born very compact in MOND cannot stay compact.

There are five bright old globular clusters orbiting around the
Fornax dwarf elliptical, 140kpc away from the Milky Way centre.
These globulars have a typical baryonic mass of $10^{4-5}\msun$
enclosed in a limiting radius of $30-50\pc$ with a half-mass
radius of $4-12\pc$ (using data compiled in Mackey \& Gilmore
2003 and Rodgers \& Roberts 1994). One of clusters, No. 1, has
a distorted profile and appears to be tidally disturbed. The
clusters No. 3,4,5 appear to have excess stars near their limiting
radii. The clusters No.1,2,5 are on the outskirts of Fornax,
between 1-1.5kpc from the Fornax centre. Our model of
relaxation-driven expansion predicts (cf. Fig.~\ref{rhage}) that
Fornax No.1 ($1\times 10^4\lsun$) should have grown to a half-mass
size of about 40pc, and more massive clusters No. 2 and No. 5
($5\times 10^4\lsun$) should grow to 12pc. The predicted sizes are
significantly greater than the observed half-light radius of 12pc
(No.1) and 6pc (No.2 and No.5).

Our model also predicts (cf. eq.~\ref{roche}) that the Roche lobe of
these Fornax clusters should have a radius of $(650-1000)\pc$ if
they are on a 140kpc orbit around the Milky Way ($5\times
10^{10}\lsun$), or a Roche lobe radius of $(100-130)\pc$ if they are
on a 1.5 kpc orbit around Fornax ($2\times 10^7\lsun$). Both of
these predictions are much greater than the observed limiting
radius of (40-50)pc for all Fornax clusters.

\subsection{Discussion}

The limiting size to which a globular can maintain internal
equilibrium are most likely limitted by the tidal force of the
Milky Way in general. This is supported by the observation of a
characteristic sudden change of outer density profile of most
globulars and transient non-spherical outer features, as
illustrated by the beautiful stream of Pal~5 found by Sloan
Digital Sky Survey, and many distant clusters, e.g., NGC5694 and
NGC1904, when deep observations have been taken along lines of
sight with least background confusion (Leon et al. 2000,
Odenkirchen et al. 2003, Grillmair et al. 1995,1996, Lehmann \&
Scholz 1997). These features are most likely due to stars leaking
from a boundary set by the Roche lobe much like the donor star in
a binary evolving off the main sequence and transfering mass to
its companion. The MOND gravity has a $\ln(r)$ infinite potential
well, and so if in isolation a globular will grow indefinitely due
to two-body relaxation.  So both two-body relaxation and violent
relaxation tend to smooth out any sharp initial boundaries. It is
hard to imagine a non-tidal mechanism to sustain sudden changes of
the density profiles.  A globular must have a finite potential
well or an outer truncation also because large amount of recycled
gas from aging and evolved stars must flow out of the tidal radii
of a globular as old globulars keep less than $1\msun$ gas inside
(Spergel 1991).

\begin{figure}
\resizebox{9cm}{!}{\includegraphics{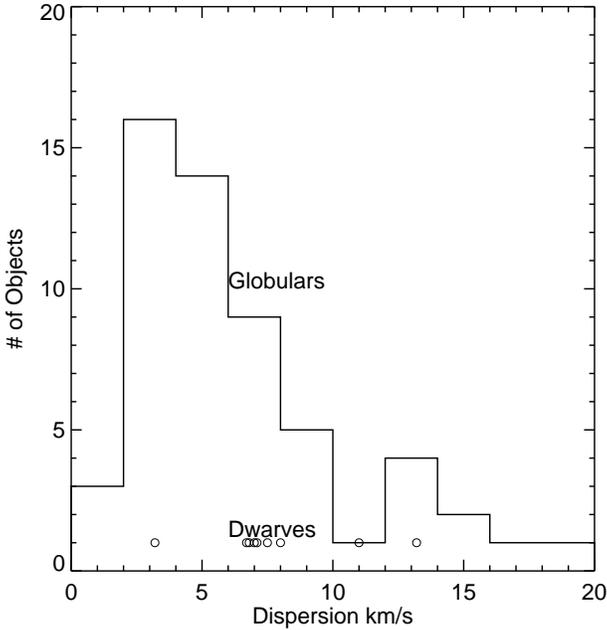}} \caption{ shows
the central line of sight velocity dispersion of Galactic globular
clusters (in histogram), and the velocity dispersion of Galactic
dwarf galaxies (circles).  Data taken from Gnedin et al. (2002)
and Mateo (1998).  }\label{dispersion}
\end{figure}

The cores of many Galactic globulars are as hot as dwarf galaxies
with central velocity dispersion $\ge
8$km/s (cf. Fig.~\ref{dispersion}), and some dwarf galaxies,
e.g., Sextans (Kleyna et al. 2004) are colder than most globulars.
An isolated object enters weak
regime if bigger than $\sqrt{GM/a_0} \sim
3.5\pc\sqrt{M/10^4\msun}$, hence dwarf galaxies, and outer
globulars like NGC2419 and Palomar clusters are all partly or
entirely in the weak gravity regime. Nature sometimes put star
clusters inside a small galaxy (as in Fornax, WLM and Sagittarius
dwarf galaxies), but never vice versa. In fact there are evidences
that some globulars are stripped out from the envelopes or carved
from the cores of dwarf galaxies by galactic tide, which explains
the similar orbits and smooth transition of baryonic mass of these
two populations of satellites. It seems that while MOND-like
gravity theories can account rotation curves of galaxy-scale
structures without fine tuning, they do not genericly come with an
important ingredient or sub-galaxy scale parameter that
distinguishes dwarfs from globulars.  Having dark matter in
dwarves but not in globulars is perhaps one way out of the dilemma
of a baryonic universe.

\begin{acknowledgments}
This proceeding was partly written during an extended visit to Beijing Observatory,
funded through the Overseas Outstanding Young Scholarship program of
the Chinese National Science Fundation.
I would like to thank the Beijing Observatory for hospitality.
I also acknowledge discussions with David Bacon, Xuelei Chen, Douglas Heggie,
Keith Horne and Andy Taylor.
\end{acknowledgments}




\end{document}